\documentclass[11pt]{article}
\usepackage{graphicx}
\usepackage{amssymb}
\usepackage{amscd}
\usepackage{mathrsfs}
\usepackage{longtable,lscape}
\usepackage{amsthm}
\usepackage{amsfonts}
\usepackage{amsmath}
\usepackage{bbm}
\usepackage{float}
\usepackage{url}
\usepackage{csquotes}
\usepackage{wrapfig}

\usepackage{caption}
\usepackage{lineno}
\usepackage[title]{appendix}
\begin{document}
\title{\bf Are Bell-tests only about local incompatibility?
}
\author{Diederik Aerts and Massimiliano Sassoli de Bianchi\vspace{0.5 cm} \\ 
\normalsize\itshape
Center Leo Apostel for Interdisciplinary Studies, \\ \itshape Brussels Free University, 1050 Brussels, Belgium\vspace{0.5 cm} \\ 
\normalsize
E-Mails: \url{diraerts@vub.ac.be}, \ \url{msassoli@vub.ac.be}
}
\date{}
\maketitle
\begin{abstract} 
\noindent The view exists that Bell-tests would only be about \emph{local} incompatibility of quantum observables and that quantum \emph{non-locality} would be an unnecessary concept in physics. In this note, we emphasize that it is not incompatibility at the \emph{local} level that is important for the violation of Bell-CHSH  inequality, but incompatibility at the \emph{non-local} level of the joint measurements. Hence, non-locality remains a necessary concept to properly interpret the outcomes of certain joint quantum measurements. 
\end{abstract} 
\medskip
{\bf Keywords:} Bell-CHSH inequality; Bell-test, contextuality; complementarity; compatibility; non-locality; non-spatiality; marginal laws, no-signaling conditions
\vspace{0.5cm}

Some authors have argued that since Bell-CHSH inequality is only violated under the condition of local incompatibility of Alice's and Bob's observables, Bell-tests should only be considered as special tests of incompatibility of said local observables, hence Bell's introduction of the very notion of non-locality would be misleading and the term ``non-locality'' should be dismissed altogether \cite{Khrennikov2017,Khrennikov2019,Khrennikov2019b,Khrennikov2020, Griffiths2020}. More precisely, addressing here the more specific point raised in \cite{Khrennikov2019}, and following Khalfin and Tsirelson's algebraic method \cite{KhalfinTsirelson1985}, we observe that by taking the square of the CHSH operator
\begin{equation}
C=A\otimes (B+B') + A'\otimes (B-B'),
\label{CHSH operator}
\end{equation}
then using the fact that the observables $A$, $A'$, $B$ and $B'$ have $\pm 1$ eigenvalues, one can write:
\begin{equation}
C^2=4\mathbb{I} +[A,A'] \otimes [B, B'].
\label{square CHSH operator}
\end{equation}
Since the inequality $C^2\leq 4\mathbb{I}$ (which implies $|\langle \psi |C|\psi\rangle| \leq 2$, for all $\psi$, which is the usual statement of the Bell-CHSH inequality; see for instance \cite{Sassoli2020} for the details) can only be violated if $[A, A'] = 0$ and/or $[B, B'] = 0$, one finds that the Bell-CHSH inequality cannot be violated if Alice's and/or Bob's observables are compatible, i.e., commute. Based on this observation, one might be tempted to conclude that Bell-tests cannot truly highlight the presence of non-locality, but only of local incompatibility (local non-commutability) of Alice's and Bob's measurements. 

The above reasoning is however incomplete, as it does not take into account the reason why local non-commutativity is necessary in the first place. To show this, let us start considering the situation where Alice's and Bob's measurements are compatible, so that we have the commutation relations $[A, A'] = 0$ and $[B, B'] = 0$. If so, one can in principle define a single measurement scheme for Alice, consisting in jointly measuring the two observables $A$ and $A'$, as well as a single measurement scheme for Bob, consisting in jointly measuring the two observables $B$ and $B'$. Let us denote ${\cal A}$ and ${\cal B}$ the observables associated with these two bigger local measurements, performed by Alice and Bob, respectively. If $A$ and $A'$ are 2-outcome observables, this means that ${\cal A}$ is associated with the 4 outcomes $(A_1,A'_1)$, $(A_2,A'_1)$, $(A_1,A'_2)$, $(A_2,A'_2)$, so that the outcome-probabilities for the two sub-measurements $A$ and $A'$ can be deduced as marginals of the outcome-probabilities of such bigger local measurement; and the same holds true for Bob's observable ${\cal B}$, associated with the 4 outcomes $(B_1,B'_1)$, $(B_2,B'_1)$, $(B_1,B'_2)$ and $(B_2,B'_2)$.

If we additionally assume that the measurement defined by jointly executing Alice's and Bob's measurements is properly described in terms of a tensor product observable ${\cal A}\otimes {\cal B}$, as is usually done in standard quantum mechanics, i.e., by the product of the two commuting observables ${\cal A}\otimes \mathbb{I}$ and $\mathbb{I}\otimes {\cal B}$, it is clear that the overall experimental situation can be described in terms of a single measurement, defined by the action of Alice and Bob jointly performing ${\cal A}$ and ${\cal B}$. Such single measurement would produce the following 16 possible outcomes:
\begin{eqnarray}
&&((A_1,A'_1),(B_1,B'_1)),\, ((A_1,A'_1),(B_2,B'_1)),\, ((A_1,A'_1),(B_1,B'_2)),\, ((A_1,A'_1),(B_2,B'_2)),\nonumber\\
&&((A_2,A'_1),(B_1,B'_1)),\, ((A_2,A'_1),(B_2,B'_1)),\, ((A_2,A'_1),(B_1,B'_2)),\, ((A_2,A'_1),(B_2,B'_2)),\nonumber\\
&&((A_1,A'_2),(B_1,B'_1)),\, ((A_1,A'_2),(B_2,B'_1)),\, ((A_1,A'_2),(B_1,B'_2)),\, ((A_1,A'_2),(B_2,B'_2)),\nonumber\\
&&((A_2,A'_2),(B_1,B'_1)),\, ((A_2,A'_2),(B_2,B'_1)),\, ((A_2,A'_2),(B_1,B'_2)),\, ((A_2,A'_2),(B_2,B'_2)).
\label{outcomes}
\end{eqnarray}
From their probabilities, one can easily deduce those of the 4 different possible joint sub-measurements, for instance the one obtained by considering sub-measurement $A$ in association with sub-measurement $B$. More precisely, the outcome-probability ${\cal P}(A_1,B_1)$, of obtaining outcome $A_1$ for $A$ and outcome $B_1$ for $B$, would be given by the sum: 
\begin{eqnarray}
{\cal P}(A_1,B_1)&=& {\cal P}((A_1,A'_1),(B_1,B'_1)) + {\cal P}((A_1,A'_1),(B_1,B'_2))\nonumber\\
&+& {\cal P}((A_1,A'_2),(B_1,B'_1)) + {\cal P}((A_1,A'_2),(B_1,B'_2)),
\label{outcomes2}
\end{eqnarray}
and similarly for the other outcome-probabilities. 

Now, the probabilities deduced from a single measurement situation can always fit into a single Kolmogorovian probability space, and therefore be represented in terms of deterministic, non-contextual hidden-variables; see for instance the representation theorem in \cite{AertsSozzo2012}. In other words, if all measurements performed by Alice and Bob are compatible, no probabilistic structure extending beyond the classical one can be revealed. This means that in order to highlight the existence of elements of reality that cannot be described by classical probability models, and therefore by classical hidden-variables theories, one needs to consider situations where an entity is not always subjected to the same experimental context, i.e., to the same measurement, however big such measurement is. Bell's work was precisely about identifying and analyzing experimental situations able to produce joint probabilities that cannot be modeled using a single Kolmogorovian probability space, in order to test whether the type of correlations identified by EPR \cite{EPR1935} did truly exist in reality. 

In the case of the Bell-CHSH inequality, 4 different measurements are required and represented by the tensor product observables $A\otimes B$, $A'\otimes B$, $A\otimes B'$ and $A'\otimes B'$. Using $A^2=A'^2=B^2=B'^2=\mathbb{I}$, one can then deduce the the 6 commutation relations: 
\begin{eqnarray}
&&[A\otimes B,A'\otimes B]= [A,A']\otimes \mathbb{I},\nonumber\\
&&[A\otimes B,A\otimes B']= \mathbb{I}\otimes [B,B'],\nonumber\\
&&[A'\otimes B,A'\otimes B']= \mathbb{I}\otimes [B,B'],\nonumber\\
&& [A\otimes B',A'\otimes B']= [A,A']\otimes \mathbb{I}, \nonumber\\
&& [A\otimes B,A'\otimes B']= [A,A']\otimes BB' + A'A\otimes [B,B'], \nonumber\\
&&[A'\otimes B,A\otimes B']= [A',A]\otimes BB' + AA' \otimes [B,B'].
\label{commutations}
\end{eqnarray}
It is clear from the above  that for the 4 observables $A\otimes B$, $A'\otimes B$, $A\otimes B'$ and $A'\otimes B'$, to describe measurements that cannot be incorporated into a single measurement scheme, we must have $[A,A']\neq 0$ and/or $[B,B']\neq 0$, i.e., Alice's and Bob's local observables must not all commute. However, this transfer of the incompatibility requirement from the non-local to the local level of the observables, is only the consequence of the fact that a specific representational choice has been a priori adopted: that of describing all joint measurements between Alice and Bob as product measurements relative to a \emph{unique} tensor product representation of the state space. But this is a very special situation, which will not necessarily apply to all experimental situations. For instance, it will certainly be invalid if in addition to the Bell-CHSH inequality also the marginal laws (also called no-signaling conditions) are violated, as observed in many experiments \cite{AdenierKhrennikov2007,DeRaedt2012,DeRaedt2013,AdenierKhrennikov2016,Bednorz2017,Kupczynski2017}.\footnote{Note that the marginal laws can be easily violated when joint measurements are performed on spatially interconnected macroscopic entities, as well as on conceptual entities that are connected through meaning; see \cite{AertsEtAl2019} and the references cited therein.}

But even when the marginal laws are obeyed, a single tensor product representation for all the observables will not work if the Bell-CHSH inequality is violated beyond Tsirelson's bound, as it is the case for, say, the ``Bertlmann wears no socks'' experiment described in \cite{AertsSassoli2019}. This is an experimental situation where the Bell-CHSH inequality  is maximally violated even though Alice's measurements $A$ and $A'$, and Bob's measurements $B$ and $B'$, are perfectly compatible at the local level. In other words, this is a situation that cannot be described by (\ref{square CHSH operator}), as the violation originates from an incompatibility which manifests at the non-local level of the joint measurements, the reason being that the correlations are created by the joint action of Alice and Bob, in a purely contextual way, i.e., the common causes at the origin of the correlations are contextually actualized. 

So, while being true that incompatibility does play a central role in the construction of a Bell-test experiment, and in the understanding of its rationale, it is incompatibility at the global level of the joint measurements that is fundamental to have, which only reduces to local incompatibility when all the entanglement can be ``pushed'' into the state of the system. This is only possible, within the standard Hilbertian formulation of quantum mechanics, if the Bell-CHSH inequality is violated below Tsirelson's bound and all marginal laws are satisfied. In more general situations, entanglement needs to be allocated also at the level of measurements, as they cannot anymore be all described as product observables relative to a same tensor product representation. This is of course a manifestation of contextuality: the possibility of using a tensor product representation for the observables describing joint measurements becomes contextual, in the sense that one needs to adopt a different isomorphism for each joint measurement, in order to allocate the entanglement resource only in the state; see \cite{AertsSozzo2014,AertsEtAl2019} 
for the details. 

To put the above differently, the incompatibility of the different joint measurements means that one cannot find a non-contextual hidden-variables representation for the observed outcome-probabilities, i.e., one cannot find common causes in the past explaining \emph{all} the correlations that are revealed by the experimental data. So, one is forced to recognize that these correlations were not all pre-existing the measurements, that some of them (or all of them) were contextually created by the latter. And this means that the common causes at the origin of the correlations are in turn genuinely contextual, where contextual means here that they are 
actualized at each run of a joint measurement in a way that depends on the type of joint measurement that is being executed \cite{AertsSassoli2019}. Thus, if it is correct to say that Bell-test experiments are about evidencing the presence of incompatibility, i.e., the fact that not all measurements can be jointly performed, this does not mean that no conclusion can be drawn about the underlying reality producing such incompatibility. Indeed, we know that Alice's and Bob's laboratories can be located at arbitrary distance in space and that despite that, their joint action can still create correlations in a way that depends on the operations they  jointly and simultaneously perform. But if their remotely performed joint actions are able to create correlations, or better, the common causes that are at their origin, how can a discussion about non-locality be avoided? We do not refer here to a notion of non-locality in the naif sense of ``something spooky traveling in space at superluminal speed,'' but in the sense of something that can operate from a (non-spatial) layer of our physical reality, not being affected by spatial distances. 

One can of course dislike the idea that our world, at its core, would be non-local, i.e., non-spatial, and certainly saying what something ``is not'' is just a first step in an investigation. The next step is about explaining what the nature of a non-spatial entity would be, and how it would relate to our spatial domain, in which it can leave traces, for instance in the form of impacts in our measuring apparatuses. To tentatively take that second step, our group in Brussels worked out in the past years a challenging  hypothesis, according to which the micro-physical entities would be endowed of a conceptual nature, similar to that of the human concepts \cite{Aertsetal2018,AertsLester2019}. Non-spatiality would then be an expression of the fact that the micro-physical entities, being essentially conceptual in nature,\footnote{Just as electromagnetic waves and acoustic waves share the same undulatory nature, but remain completely distinct entities, the same would apply to human conceptual entities and the microscopic entities: they would share a same conceptual nature, while remaining completely distinct entities.} can be in more or less abstract states, with the less abstract ones (i.e., the more concrete ones) being precisely those associated with the condition of ``being in space.'' 

A few additional remarks are in order, to better understand what a Bell-test can tell us about the reality of a composite micro-system. We mentioned that a single tensor product representation can only be used when the Bell-CHSH inequality is violated below Tsirelson's bound and all marginal laws are satisfied. This should lead one to reflect on the practice of mathematically representing joint measurements by product operators, and then have superpositions of product states to describe the presence of entanglement. 

If we accept the idea that density operators can also represent genuine states, then the situation is not in conflict with the general physical principle saying that a composite system exists, and therefore is in a well-defined state, only if its sub-systems also exist, and therefore are also in well-defined states \cite{AertsSassoli2016}. On the other hand, if we believe that only vector-states (i.e., pure states) can represent genuine states, then the choice of representing the state of a composite system as a superposition of product states, not allowing then to attach vector-states to the sub-systems, becomes questionable, as physically unintelligible (how could the sub-systems exist if they are not in well-defined states?). 

A different possibility would be to drop the requirement of describing entanglement as a property of the state, i.e., to use a product state to describe the state of the system and then non-product operators to describe the joint measurements. In other words, when confronted with a violation of the Bell-CHSH inequality, there is no a priori requisite to mathematically model the experimental situation in terms of entangled states: entangled (non-product) measurements can also be used, and in fact must be used if the marginal laws are also disobeyed. 

But even this situation is not full satisfactory, as is clear that when a system evolves, a product state will generally transform into a non-product state, hence the interpretational problem remains and the use of density operators to describe genuine states seems to be inescapable \cite{AertsSassoli2016}. 

The above issue can be better understood if one observes, as one of us did many decades ago, that in the Hilbert space formalism separated entities cannot be consistently modeled. The reason for this is that the Hilbertian formalism is, structurally speaking, too specific, as it satisfies an axiom called the `covering law' \cite{Aerts1980,Aerts1982Found,Aerts1984a,Sassoli2020}. This means that non-locality, which in ultimate analysis means non-separability, is intrinsically part of the quantum formalism, so much so that locality/separability cannot be even properly expressed within it (not for a lack of states, but for a lack of properties). In other words, the very decision of using a Hilbert space to model a physical system already and unavoidably introduces non-locality/non-separability in its description. 

Note that the Bell-CHSH inequality, being expressed only in terms of probabilities, is independent of the mathematical formalism used to model an experimental situation. Also, Bell was primarily interested in separability, and not whether the probability structure of a composite system would be Kolmogorovian or non-Kolmogorovian, and in particular if the probabilities of the different joint measurements could be described as the marginals of a unique joint probability distribution. In particular, Bell wasn't focused on the marginal laws being satisfied or not. This question only came about later, with the analysis of Fine \cite{Fine1982}, Pitowsky \cite{p1989} and more recently Dzhafarov and Kujala \cite{dk2014}. 

Note also that the special attention placed on the marginal laws only resulted from their interpretation as `no-signaling conditions' \cite{GhirardiEtAl1980,Ballentine1987}. Indeed, it has been argued that if violated they could be used to achieve faster than light communication. However, a more attentive analysis of the situation shows that this is not necessarily the case, for at least two reasons: it is not clear what are the times involved, in order to handle a large enough statistical ensemble of identically prepared systems, and if, when they are properly accounted for, they would still allow for an effective supraluminal communication. Also, and more importantly, the existence of correlations separated by space-like intervals does not per se imply that they result from underlying phenomena propagating in space faster than the speed of light, as the numerous models investigated by our group have clearly shown; see for instance the discussions in \cite{aabgssv2018b,AertsEtAl2019,Sassoli2020,AertsSassoli2019}

Our digression on marginal laws allows us to ``close the circle'' of our analysis. We mentioned that the description of all the observables associated with the different joint measurements in a Bell-test experiment, as product observables relative to a single tensor product representation, is what creates the illusion that Bell-tests would only be about local incompatibility. There is however no a priori physical justification to believe that such a peculiar representation would correspond to the general case. Certainly, it is not forced upon us by the available empirical data, which in fact tell us a rather different story, considering that the marginal laws are typically violated. 

The proof of the marginal laws also relies on the existence of a single tensor product representation, which however, again, is not imposed by the quantum formalism (see in particular the discussion in \cite{Kennedy1995}), hence should be justified by the data. 

To conclude, Bell-tests are not just about local incompatibility: taking into account the available data, and until proven to the contrary, they also are about non-local incompatibility, hence about non-locality. Furthermore, the marginal laws are not genuine no-signaling conditions, as their violation does not necessarily imply a faster than light propagation of signals in space, and they also result of the peculiar choice (which requires a physical justification) of representing all joint measurements in a given experimental situation as product measurements relative to a unique tensor product decomposition.


\begin{thebibliography}{}
\setlength{\itemsep}{-.5mm}

\bibitem{Khrennikov2017} Khrennikov, A. (2017). Bohr against Bell: complementarity versus nonlocality, Open Phys. 15: 734--738.
\bibitem{Khrennikov2019} Khrennikov, A. (2019). Violation of the Bell’s type inequalities as a local expression of incompatibility, J. Phys.: Conf. Ser. 1275, 012018.
\bibitem{Khrennikov2019b} Khrennikov, A. (2019). Get Rid of Nonlocality from Quantum Physics, Entropy 2019, 21, 806.
\bibitem{Khrennikov2020} Khrennikov, A. (2020). Two faced Janus of quantum nonlocality, Entropy 22, 303.
\bibitem{Griffiths2020} Griffiths, R. B. (2020).  Nonlocality Claims are Inconsistent with Hilbert Space Quantum Mechanics, Phys. Rev. A 101, 022117.
\bibitem{KhalfinTsirelson1985} Khalfin, L. A. \& Tsirelson, B. S. (1985). Quantum and quasi-classical analogs of Bell inequalities.
In: Symposium on the Foundations of Modern Physics 1985 (ed. Lahti et al.; World Sci. Publ.), 441--460.
\bibitem{Sassoli2020} Sassoli de Bianchi, M. (2020). Violation of CHSH inequality and marginal laws in mixed sequential measurements with order effects. Soft Computing 24, 10231--10238.
\bibitem{AertsSozzo2012} Aerts, D. \& Sozzo, S. (2012a). Entanglement of conceptual entities in Quantum Model Theory (QMod). {\it Quantum Interaction. Lecture Notes in Computer Science 7620}, 114--125.
\bibitem{EPR1935} Einstein, A. Podolsky, B. \& Rosen, N. (1935). Can Quantum-Mechanical Description of Physical Reality Be Considered Complete?, Phys. Rev. 47,  777--780.
\bibitem{AdenierKhrennikov2007} Adenier, G. \& Khrennikov, A. (2007). Is the fair sampling assumption supported by EPR experiments?, J. Phys. B: Atomic, Molecular and Optical Physics 40, 131--141.
\bibitem{DeRaedt2012} De Raedt, H., Michielsen, K. \& Jin, F. (2012). Einstein-Podolsky-Rosen-Bohm laboratory experiments: Data analysis and simulation, AIP Conf. Proc. 1424, 55--66.
\bibitem{DeRaedt2013} De Raedt H., Jin, F. \& Michielsen, K. (2013). Data analysis of Einstein-Podolsky-Rosen-Bohm laboratory experiments. Proc. of SPIE 8832, The Nature of Light: What are Photons? V, 88321N.
\bibitem{AdenierKhrennikov2016} Adenier, G. \& Khrennikov, A. (2017). Test of the no-signaling principle in the Hensen loophole-free CHSH experiment, Fortschritte der Physik (Progress in Physics) 65, 1600096. 
\bibitem{Bednorz2017} Bednorz A. (2017). Analysis of assumptions of recent tests of local realism, Phys. Rev. A 95, 042118.
\bibitem{Kupczynski2017} Kupczynski, M. (2017). Is Einsteinian no-signalling violated in Bell tests?, Open Phys. 15, 739--753.
\bibitem{AertsSozzo2014} Aerts, D. \& Sozzo, S. (2014). Quantum Entanglement in Concept Combinations. Int. J. Theor. Phys. 53, 3587--3603.
\bibitem{AertsEtAl2019} Aerts, D., Aerts Argu\"elles, J., Beltran, L., Geriente, S., Sassoli de Bianchi, M., Sozzo, S \& Veloz, T. (2019). Quantum entanglement in physical and cognitive systems: a conceptual analysis and a general representation. European Physical Journal Plus 134: 493.
\bibitem{AertsSassoli2019} Aerts, D. \& Sassoli de Bianchi, M. (2019). When Bertlmann wears no socks: contextual common causes as an explanation for quantum correlations. arXiv:1912.07596 [quant-ph]. 
\bibitem{Aertsetal2018} Aerts, D., Sassoli de Bianchi, M., Sozzo, S. \& Veloz, M. (2020). On the Conceptuality interpretation of Quantum and Relativity Theories. Foundations of Science 25, 5--54.
\bibitem{AertsLester2019} Aerts, D. \& Beltran, L. (2019). Quantum Structure in Cognition: Human Language as a Boson Gas of Entangled Words. Foundations of Science. https://doi.org/10.1007/s10699-019-09633-4.
\bibitem{AertsSassoli2016} Aerts, D. \& Sassoli de Bianchi, M. (2016). The extended Bloch representation of quantum mechanics: Explaining superposition, interference, and entanglement,  Journal of Mathematical Physics 57, 122110.
\bibitem{Aerts1980} Aerts, D. (1980).  Why is it impossible in quantum mechanics to describe two or more separated entities. Bulletin de l'Academie royale de Belgique, Classes des Sciences 66, 705--714.
\bibitem{Aerts1982Found} Aerts, D. (1982). Description of many physical entities without the paradoxes encountered in quantum mechanics, Found. Phys.  12, 1131--1170.
\bibitem{Aerts1984a} Aerts, D. (1984). The missing elements of reality in the description of quantum mechanics of the EPR paradox situation, Helv. Phys. Acta 57, 421--428.
\bibitem{Sassoli2019} Sassoli de Bianchi, M. (2019). On Aerts' overlooked solution to the EPR paradox.  In: Probing the Meaning of Quantum Mechanics. Information, Contextuality, Relationalism and Entanglement. D. Aerts, M.L. Dalla Chiara, C. de Ronde \& D. Krause (eds.) World Scientific, pp. 185--201.
\bibitem{Fine1982} Fine, A. (1982). Joint distributions, quantum correlations, and commuting observables. Journal of Mathematical Physics 23, 1306--1310.
\bibitem{p1989} Pitowsky, I. (1989). {\it Quantum Probability, Quantum Logic}. Lecture Notes in Physics vol. {\bf 321}.  Berlin: Springer.
\bibitem{dk2014} Dzhafarov, E. N., \& Kujala, J. V. (2017). Fortschr. Phys. 65, 1600040.
\bibitem{Ballentine1987} Ballentine, L. E. \& Jarrett, J. P. (1987). Bell's theorem: Does quantum mechanics contradict relativity? American Journal of Physics 55, 696--701; doi: 10.1119/1.15059.
\bibitem{GhirardiEtAl1980} Ghirardi, G. C., Rimini, A. \& Weber, T. (1980). A General Argument against Superluminal Transmission through the Quantum Mechanical Measurement Process. Lett. Nuovo Cimento 27, 293--298.
\bibitem{aabgssv2018b} Aerts, D., Aerts Argu\"elles, J., Beltran, L., Geriente, S., Sassoli de Bianchi, M., Sozzo, S \& Veloz, T. (2018). Spin and wind directions II: A Bell State quantum model. Foundations of Science 23, 337--365.
\bibitem{Kennedy1995} Kennedy, J. B. (1995). On the Empirical Foundations of the Quantum No-Signalling Proofs, Philosophy of Science 62, 543--560.

\end{thebibliography}
\end{document}